\pdfoutput=1
\RequirePackage{ifpdf}
\ifpdf %We are running pdfTeX in pdf mode
\documentclass[pdftex]{sigma}
\else
\documentclass{sigma}
\fi

\numberwithin{equation}{section}
\begin{document}

\allowdisplaybreaks

\renewcommand{\thefootnote}{$\star$}

\renewcommand{\PaperNumber}{104}

\FirstPageHeading

\ShortArticleName{Maximal Energy Scale in Photon Gas Thermodynamics and	Construction of Path Integral}

\ArticleName{Ef\/fects of a~Maximal Energy Scale\\ in Thermodynamics for Photon Gas\\ and Construction of Path
Integral\footnote{This paper is a~contribution to the Special Issue on Deformations of Space-Time and its Symmetries.
The full collection is available at \href{http://www.emis.de/journals/SIGMA/space-time.html}
{http://www.emis.de/journals/SIGMA/space-time.html}}}

\Author{Sudipta DAS, Souvik PRAMANIK and Subir GHOSH}
\AuthorNameForHeading{S.~Das, S.~Pramanik and S.~Ghosh}

\Address{Physics and Applied Mathematics Unit, Indian Statistical Institute,\\
203 B.T.~Road, Kolkata 700108, India}

\Email{\href{mailto:sudipta_jumaths@yahoo.co.in}{sudipta\_jumaths@yahoo.co.in},
\href{mailto:souvick.in@gmail.com}{souvick.in@gmail.com}, \href{mailto:sghosh@isical.ac.in}{sghosh@isical.ac.in}}

\ArticleDates{Received April 13, 2014, in f\/inal form October 25, 2014; Published online November 07, 2014}

\Abstract{In this article, we discuss some well-known theoretical models where an observer-independent energy scale or
a~length scale is present.
The presence of this invariant scale necessarily deforms the Lorentz symmetry.
We study dif\/ferent aspects and features of such theories about how modif\/ications arise due to this cutof\/f scale.
First we study the formulation of energy-momentum tensor for a~perfect f\/luid in doubly special relativity (DSR), where
an energy scale is present.
Then we go on to study modif\/ications in thermodynamic pro\-per\-ties of photon gas in DSR.
Finally we discuss some models with generalized uncertainty principle (GUP).}


\Classification{83A05; 82D05; 70H45}

\renewcommand{\thefootnote}{\arabic{footnote}}
\setcounter{footnote}{0}

\section{Introduction}

Quantum gravity ideas naturally suggest a~smallest (but f\/inite) observer independent length scale~$l$, or a~f\/inite
upper bound of energy~$\kappa$, which can avoid the paradoxical situation of spontaneous creation of black holes inside
a~very small region.
It is quite suggestive to consider this length scale to be the Planck length $l_{\rm P}$ itself and the energy upper
bound~$\kappa$ to be the Planck energy.
From another point of view, in the proposed quantum theories of gravity such as loop quantum gravity, the Planck length
denotes a~threshold below which the classical picture of smooth spacetime geometry gives way to a~discrete quantum
geometry.
This suggests that the Planck length plays a~role analogous to the atomic spacing in condensed matter physics.
Below that length there is no concept of a~smooth metric.
Thus the quantities involving the metric, such as the usual mass-shell condition in special relativity (SR)
\begin{gather*}
E^2 = p^2 + m^2
\end{gather*}
receive corrections of order of the Planck length such as~\cite{piran}
\begin{gather}
E^2 = p^2 + m^2 + l_{\rm P} E^3 + \cdots,
\label{moddisp}
\end{gather}
where $l_P$ is of the order of the Planck length.

However the idea of such an observer-independent length scale immediately raises a~contradiction with the principles of SR theory.
As lengths are not invariant under Lorentz transformations in SR, so one observer's threshold length scale will be
perceived to be dif\/ferent than another's, which directly contradicts the idea of an observer-independent length scale,
such as the Planck length.

\looseness=-1
A~modif\/ied energy-momentum relationship such as \eqref{moddisp} generally induces an energy dependent speed of light. In a theory with a varying speed of light, it may be the case that the speed of light was greater in the very early universe, when the energy was high enough~\cite{ncvsl}. This could be an alternative to the horizon problem which still cannot be fully explained by inf\/lation~\cite{am, moffat93}. It may also lead to corrections to the predictions of inf\/lationary cosmology, which can be verif\/ied through future CMB observations. In~\cite{laura}, it has been even argued that the still unaccounted dark energy could be mimicked by these models with modif\/ied dispersion relations. The explanation lies in the fact that the missing dark energy can be trapped by very high momentum and low-frequency quanta from transPlanckian regime, frozen at present epoch~\cite{laura}. But here lies the same problem: such a modif\/ication in the dispersion relation contradicts the Lorentz transformation laws of~SR. In SR, energy and momentum transform according to the Lorentz transformations and this Lorentz invariance is considered to be a fundamental principle in all the physical theories. Thus it is a big reason to worry that to incorporate an observer independent length (or energy) scale or to modify the canonical dispersion relation, it would break Lorentz invariance.

These paradoxes may be resolved if the Lorentz transformations could be modif\/ied so as to preserve an energy or momentum scale. In~\cite{doub, nv, leejoao}, the authors have shown that it is possible to build models where the laws of transformation of energy and momenta between dif\/ferent inertial observers are modif\/ied while keeping the principle of relativity for inertial observers intact. This can be achieved by adding nonlinear terms to the Lorentz transformations acting on momentum space. As a result, all observers agree to the presence of an invariant energy or momentum. The idea of a smooth spacetime background breaks down above this observer-independent energy threshold. In these models, one has to replace the quadratic invariant by a nonlinear invariant, thus producing a modif\/ied dispersion relation. As said earlier, in these theories, there are two invariant quantities,~$c$, the velocity of light and~$\kappa$, an upper limit of energy. As there are two observer-independent invariant quantities, this theory is named ``doubly (or deformed) special relativity'' (DSR). This DSR theory possesses the following features\footnote{For a~detail discussion on relative locality, please see~\cite{relocal}.}:
\begin{itemize}\itemsep=-1pt
\item[(i)] The relativity of inertial frames, as proposed by Galileo, Newton and Einstein, is preserved in DSR.
\item[(ii)] There is an {\it {invariant energy scale}}~$\kappa$, which is of the order of the Planck scale.
\item[(iii)] In general, DSR theory exhibits a~varying speed of light at high energies.
\item[(iv)] For this DSR theories, the notion of absolute locality
should be
replaced by relative locality as due to the
presence of an energy-dependent metric, dif\/ferent observers live in dif\/ferent spacetime.
\end{itemize}

It is possible to achieve all of these conditions through a~nonlinear action of the usual Lorentz group on the physical states of the theory. This nonlinear action immediately invokes some novel features into DSR theory:

(i) If one adds momenta and energy linearly, as we normally do in physics, the conservation of momentum becomes
inconsistent with this new nonlinear action of the lorentz group on momentum space.
Thus for energy and momentum to be conserved, the addition rules become nonlinear.
This issue of nonlinearity is particularly important for multi-particle systems in DSR.
This can be demonstrated as following: for multi-particle systems, using linear addition rule for energy/momentum leads
to a~paradox, known as ``soccer ball problem''.
The problem lies in the fact that if we apply linear addition rule for momenta/energies of many sub-Planck energy particles
then we may end up with a~multi-particle state, such as a~soccer ball
whose total energy becomes greater than the Planck energy,
which is forbidden in the DSR theory.
As we will see later, one has to apply nonlinear addition rules for energy/momentum in DSR framework, which can resolve
this paradox.
For further discussion about the ``soccer ball problem''
see~\cite{maghoss, hossrev}.

(ii) For DSR theory, spacetime
coordinates no longer commute, thus inducing a~noncom\-mu\-ta\-tive spacetime background.

Particle dynamics in DSR framework has been studied, which has revealed many unusual features~\cite{kowal, mig}.
Some f\/ield theoretic models in DSR spacetime have been attempted~\cite{fil}.
On the other hand, thermodynamics of bosons and fermions with a~modif\/ied dispersion relation and its cosmological and
astrophysical implications has been studied in~\cite{bertolami, magcos}.
In~\cite{rainbow}, authors have introduced a~procedure to incorporate gravity into DSR framework and cosmological
ef\/fects of DSR has been studied in~\cite{lin}.
We consider a~particular DSR model (for details please see the next section of this article) with~$\kappa$-deformed
Minkowski spacetime background ($\kappa$ being the quantum gravity induced noncommutative parameter), which is indeed
a~noncommutative geometry~\cite{kappa3}.
For this DSR model, the well known dispersion relation (or mass-shell condition) for a~particle
\begin{gather*}
\epsilon^2 - p^2 = m^2
\end{gather*}
has to be
modif\/ied as
\begin{gather}
\epsilon^2 - p^2 = m^2 \left(1 - \frac{\epsilon}{\kappa}\right)^2.
\label{mdr}
\end{gather}
Here $\epsilon = p^0$ and~$p$ are respectively the energy and the magnitude of the three-momentum of the particle,~$m$
is the mass of the particle and we have taken $c=1$.
With this model~\eqref{mdr}, we derive the DSR covariant energy-momentum tensor
for perfect f\/luid.
We also study the modif\/ications in thermodynamic properties of photon gas due to the presence of an upper bound of
energy,~$\kappa$, for this particular DSR model.

Another interesting idea is the generalized uncertainty principle (GUP) where the usual Heisenberg uncertainty relation
is modif\/ied as a~consequence of a~length scale presented
in the theory~\cite{gup1, gup5, gup3, gup4, gup2}.
(GUP)~\cite{ven} naturally encodes the idea of existence of a~minimum measurable length through modif\/ications in the
Poisson brackets of position~$x$ and momentum~$p$.
Indeed, one should start with the relation between the momentum and the pseudo-momentum for a~consistent deformed
algebra~\cite{hoss1}.
The Jacobi identities are then automatically fulf\/illed.
GUP has created a~lot of interest in the f\/ields like black hole thermodynamics, cosmology
and other related areas~\cite{gup10, gup9, gup7, gup6, gup8}.
In this article, we have discussed the formulation of particle Lagrangian in GUP in a~covariant manner.

It is noteworthy to mention that all the models we have studied in our works possess very rich constraint structure.
To study dynamics of these models, we use the elegant scheme of Dirac constraint analysis in Hamiltonian
framework~\cite{dirac, teit}.
Here we discuss Dirac's method of constraint analysis in brief.
In Dirac's method, from a~given Lagrangian, one starts by computing the conjugate momentum $p=\frac{\partial L}{\partial
\dot q}$ of a~generic variable~$q$ and identif\/ies the relations that do not contain time derivatives as (Hamiltonian)
constraints.
New constraints can also be generated from demanding time persistence of the f\/irst set of constraints.
Once the full set of constraints is obtained, a~constraint is classif\/ied as f\/irst class constraint (FCC) when it
commutes with all other constraints (modulo constraint) and the set of constraints which do not commute are called
second class constraints (SCC).
Presence of constraints indicates a~redundance of degrees of freedom (d.o.f.) so that not all the d.o.f.s
are independent.
FCCs present in a~theory signal gauge invariance.
The FCCs and SCCs should be
treated in essentially dif\/ferent ways.
There are two ways to deal with FCCs:
(i) either one can keep all the d.o.f.s
and impose the FCCs by restricting the set
of physical states to those satisfying $({\rm FCC})\,|\, {\rm  state} \rangle = 0$;
(ii) or one can choose additional constraints (one
each for one FCC), known as gauge f\/ixing conditions so that these, together with the FCCs turn in to an SCC set.
Now, for SCCs, a~similar relation as above, $({\rm SCC})\,|\, {\rm state} \rangle = 0$ cannot be implemented consistently and one
needs to replace the Poisson brackets by Dirac brackets to properly incorporate the SCCs.
If $(\{\psi_{\rho}^{i}, \psi_{\sigma}^{j}\}^{-1})$ is the $(i j)$-th element of the inverse constraint matrix where
$\psi^{i}(q,p)$ is a~set of SCCs, then the Dirac bracket between two generic variables $\{A(q,p), B(q,p)\}_{\rm DB}$ is given~by
\begin{gather*}
\{A, B\}_{\rm DB} = \{A, B\} - \{A, \psi_{\rho}^{i}\} \big(\{\psi_{\rho}^{i}, \psi_{\sigma}^{j}\}^{-1}\big) \{\psi_{\sigma}^{j},B\},
\end{gather*}
where $\{\,,\,\}$ denotes Poisson brackets.
Subsequently, one can quantize the theory by promoting these Dirac brackets to quantum commutators.
It should be pointed out that the noncommutative
algebras appearing in our models eventually emerge from the Dirac
brackets between the corresponding phase space variables.
In this Hamiltonian framework, the SCCs~$\psi_{\rho}^{i}$ are considered to be ``strongly'' zero since they commute with
any generic variable~$A$: $\{A, \psi_{\rho}^{i}\}_{\rm DB} = \{\psi_{\rho}^{i}, A\}_{\rm DB} = 0$, which implies a~redundance in
the number of d.o.f.s.
Hence, to understand the ef\/fect of constraints we note that the presence of one FCC together with its gauge f\/ixing
constraint can remove two d.o.f.s
from the phase space whereas one SCC can remove only one d.o.f.\
from the phase space, respectively (for details regarding Dirac's constraint analysis, please see~\cite{dirac, teit}).

This article is organized as follows: in the Sections~\ref{Section2},~\ref{Section3} and~\ref{Section4}
we discuss doubly special relativity (DSR) models where an observer-independent energy scale is present.
We explicitly show that this scale induces noncommutative spacetime background along with deforming the Lorentz
symmetry.
Treating perfect f\/luid as a~multi-particle system, we derive an expression for the energy-momentum tensor for this
perfect f\/luid in DSR.
We also study thermodynamic properties of photon gas in this DSR framework where modif\/ications are induced by the
invariant energy scale present in the theory.

In Sections~\ref{Section5} and~\ref{Section6}, we discuss some GUP induced models.
In Section~\ref{Section5}, we derive a~free particle
GUP Lagrangian in covariant manner as well as derive a~Lagrangian in presence of
an external electromagnetic f\/ield where the usual equations of motion are modif\/ied by the noncommutative parameter
present in the theory.
In Section~\ref{Section6}, we consider a~GUP Hamiltonian and consequently derive its corresponding kernel following Feynman's path
integral approach.
Finally we summarize and conclude in Section~\ref{Section7}.

\section{Deformations in Lorentz symmetry:\\ noncommutative spacetime}\label{Section2}

Here we discuss about a~well-known DSR model, known as the~$\kappa$-Minkowski spacetime.
As mentioned earlier, this~$\kappa$-Minkowski model possesses an invariant energy scale~$\kappa$.
Due to the presence of this scale, the spacetime in this~$\kappa$-Minkowski DSR model becomes noncommutative.
This noncommutativity can be explicitly seen through the underlying phase space algebra which is written in a~covariant
form,
\begin{gather*}
\{x_\mu,x_\nu \}=\frac{1}{\kappa}(x_\mu \eta_{\nu}-x_\nu\eta_{\mu}),
\qquad
\{x_{\mu},p_{\nu}\}=-g_{\mu\nu}+\frac{1}{\kappa}\eta_{\mu}p_{\nu},
\qquad
\{p_{\mu},p_{\nu}\}=0,
\end{gather*}
where $\eta_0=1$, $\eta_i=0$.
This algebra appeared in~\cite{granik} and partially in~\cite{mig}.
Detailed studies of similar types of algebra are provided in~\cite{kowal}.

It has been pointed out by Amelino-Camelia~\cite{nv}
that there is a~connection between the appearance of an observer
independent scale and the presence of nonlinearity in the corresponding spacetime transformations.
Recall that Galilean transformations are completely linear and there are no observer independent para\-me\-ters in
Galilean/Newtonian relativity.
With Einstein re\-la\-tivity one f\/inds an observer independent scale, the velocity of light, as well as a~nonlinear
relation in the velocity addition theorem.
In DSR one introduces another observer independent parameter, an energy upper bound~$\kappa$, and ushers another level
of nonlinearity in which the Lorentz transformation laws become nonlinear.
These generalized Lorentz transformation rules, referred to here as DSR Lorentz transformation, are derivable from basic
DSR ideas~\cite{nv} or in a~more systematic way, from integrating small DSR transformations in a~NC spacetime scheme~\cite{br, sg}.
Another elegant way of derivation is to interpret DSR laws as a~nonlinear realization of SR laws~\cite{visser, ms}
where one can directly exploit the nonlinear map and its inverse, that connects DSR to SR and vice-versa.
It should be pointed out that even though there exists an explicit map between SR and DSR variables, the two theories
will not lead to the same physics (in particular upon quantization), due to the essential nonlinearity involved in the map.
Also, one can equivalently say that this map or transformation is not canonical since it changes the Poisson bracket
structure in a~non-trivial way.
According to DSR the physical degrees of freedom live in a~non-canonical phase space and the canonically mapped phase space
is to be
used only as a~convenient intermediate step.
Obviously, to accomplish this, one needs the explicit expression for the map which can be constructed by a~motivated
guess~\cite{visser, ms} or constructed as a~form of Darboux map~\cite{sg}.

We are working in the DSR model of Magueijo and Smolin~\cite{ms}.
Let us start with the all important map~\cite{sg, visser, ms}
\begin{gather*}
F(X^{\mu})\rightarrow x^{\mu},
\qquad
F^{-1}(x^{\mu})\rightarrow X^{\mu},
%\label{1}
\end{gather*}
which in explicit form reads
\begin{gather*}
F(X^\mu) = x^{\mu}\left(1-\frac{p^0}{\kappa}\right) = x^{\mu}\left(1-\frac{(n p)}{\kappa}\right),
\\
F^{-1}(x^{\mu}) = X^{\mu}\left(1+\frac{P^0}{\kappa}\right) = X^{\mu}\left(1+\frac{(n P)}{\kappa}\right),
\\
F(P^\mu) = \frac{p^{\mu}}{\left(1-\frac{p^0}{\kappa}\right)} = \frac{p^{\mu}}{\left(1-\frac{(n p)}{\kappa}\right)},
\\
F^{-1}(p^{\mu}) = \frac{P^{\mu}}{\left(1+\frac{P^0}{\kappa}\right)} = \frac{P^{\mu}}{\left(1+\frac{(n P)}{\kappa}\right)},
\end{gather*}
where $n_{\mu}=(1, 0, 0, 0)$ is introduced to express the map in a~covariant way and we use the notation $a_\mu
b^\mu=(ab)$, $(n p)=p^0$, $(n P)=P^0$.
Note that upper case and lower case letters refer to (unphysical) canonical SR variables and (physical) DSR variables
respectively.
Using canonical Poisson brackets it is straightforward to generate the noncommutative phase space algebra of DSR
variables.

To derive the generalized DSR Lorentz transformations ($L_{\rm DSR}$), one starts with the familiar (linear) SR Lorentz
transformations ($L_{\rm SR}$) and then the nonlinear $L_{\rm DSR}$ can be obtained by following mechanism
\begin{gather*}
x'^\mu = L_{\rm DSR} (x^\mu) = F \circ L_{\rm SR} \circ F^{-1} (x^\mu),
\qquad
p'^\mu = L_{\rm DSR} (p^\mu) = F \circ L_{\rm SR} \circ F^{-1} (p^\mu).
\end{gather*}
In explicit form this reads as
\begin{gather*}
x'^0 = \gamma \alpha \big(x^0 - v x^1\big),
\qquad
p'^0 = \frac{\gamma}{\alpha}\big(p^0 - v p^1\big),
\end{gather*}
where $\gamma = \frac{1}{\sqrt{1-v^2}}$ and the boost is along $X^1$ direction with velocity $v^i = (v, 0, 0)$ and
$\alpha = 1 + \frac{1}{\kappa} ((\gamma-1)P^0-\gamma vP^{1})$.
Similarly for $\mu=1$, we have the following expressions
\begin{gather*}
x'^1 = \gamma \alpha (x^1 - v x^0),
\qquad
p'^1 = \frac{\gamma}{\alpha}\big(p^1 - v p^0\big).
\end{gather*}
{\it It is important to realize that, in the present formulation, noncommutative effects enter through these generalized
$($nonlinear$)$ transformation rules.}

Note that, in contrast to SR laws, components of $x^\mu$,~$p^\mu$ {\it{transverse}} to the frame velocity~$v$ are also
af\/fected in DSR to
\begin{gather*}
x'^i = \alpha x^i,
\qquad
p'^i= \frac{p^i}{\alpha},
\qquad
i=2,3.
\end{gather*}
There are two phase space quantities, invariant under DSR Lorentz transformation
\begin{gather*}
 \eta_{\mu \nu} p^\mu p^\nu / \big(1-p^0/\kappa\big)^2 \qquad \text{and} \qquad \eta_{\mu \nu} d x^\mu d x^\nu \big(1-p^0/\kappa\big)^2
 \end{gather*}
with
$\eta_{\mu\nu}=\operatorname{diag}(-1, 1, 1, 1)$.
Writing the former as
\begin{gather*}
m^2 = \eta_{\mu \nu} p^\mu p^\nu / \big(1-p^0/\kappa\big)^2
\end{gather*}
yields the well-known Magueijo--Smolin dispersion relation.
We interpret the latter invariant to provide an ef\/fective metric $\tilde{\eta}_{\mu\nu}$ for DSR
\begin{gather}
d \tau^2 = \tilde{\eta}_{\mu \nu} d x^\mu d x^\nu = \big(1 - p^0/\kappa\big)^2\eta_{\mu \nu} d x^\mu d x^\nu.
\label{ncm}
\end{gather}
From the expression of~$\alpha$, it is clear that in the limit $\kappa \rightarrow \infty$, $\alpha \rightarrow 1$ and
all the DSR results coincide with the usual expressions in SR.

\section[Fluid dynamics in~$\kappa$-Minkowski spacetime]{Fluid dynamics in~$\boldsymbol{\kappa}$-Minkowski spacetime}\label{Section3}

In this section our aim is to construct the energy-momentum tensor (EMT) of a~perfect f\/luid, that will be covariant in
the DSR framework.
Indeed, this will f\/it nicely in our future programme of pursuing a~DSR based cosmology.

\subsection{Fluid in SR theory}

A perfect f\/luid can be considered as a~system of non-interacting structureless point particles, experiencing only
spatially localized interactions among themselves.
The energy momentum tensor (EMT) for this perfect f\/luid in the rest frame is of the form~\cite{wein}
\begin{gather}
T^{\mu\nu} = \sum\limits_{i} \frac{P^{\mu}_{i}P^{\nu}_{i}}{P^0_{i}}\delta^{3}\left(X - X_{i}\right),
\label{lt1}
\end{gather}
where $P^{\mu}_{i}$ is the energy-momentum four-vector associated with the~$i$-th particle located at $X_i$.
Once again in the comoving frame it will reduce to the diagonal form
\begin{gather}
\tilde{T}^{ii} = P = \frac{1}{3} \sum\limits_{i}\frac{{\bf P}_{i}^{2}}{P^0_{i}}\delta^{3} (X - X_i ),
\nonumber
\\
\tilde{T}^{00} = D = \sum\limits_{i} P^0_i\delta^{3} (X - X_i  ),
\qquad
\tilde{T}^{i0} = \tilde{T}^{0i} = 0.
\label{lD}
\end{gather}
In the above relations $P^0_i$ stands for the energy of the~$i$-th f\/luid particle.
The thermodynamic quantities~$P$ and~$D$ represent pressure and energy density of the f\/luid.
The particle number density is naturally def\/ined as
\begin{gather}
N = \sum\limits_{i}\delta^{3} (X - X_i ).
\label{lN}
\end{gather}
The Lorentz transformation equation for $T^{\mu \nu}$ is given~by
\begin{gather*}
T^{\mu \nu}=L_{\rm SR}\big(\tilde{T}^{\mu \nu}\big)=\Lambda^\mu_{\alpha} \Lambda^\nu_{\beta} \tilde{T}^{\alpha \beta},
\end{gather*}
where~$\Lambda$ is the Lorentz transformation matrix.
For $\mu = \nu = 0$ we have
\begin{gather*}
T^{00} = \big(\Lambda^0_{0}\big)^2\tilde{T}^{00} + \big(\Lambda^0_{i}\big)^2\tilde{T}^{ii} = \gamma^2\tilde{T}^{00} + \gamma^2
v^2\tilde{T}^{11}.
\end{gather*}
The above set of equations can be integrated into a~single SR covariant tensor
\begin{gather}
T^{\mu\nu}= (P + D)U^{\mu}U^{\nu} + P\eta^{\mu\nu},
\label{lctf}
\end{gather}
where the velocity $4$-vector $U^\mu$ is def\/ined as $U^0=\gamma$, $U^i=\gamma v^i$ with $U^\mu U_\mu=-1$.

\subsection{Fluid in DSR theory}

In order to derive the expression for the DSR covariant EMT ($t^{\mu\nu}$) we shall exploit the same approach as in case of SR EMT.
Spatial rotational invariance remains intact in DSR allowing us to postulate a~similar diagonal form for DSR EMT in the
comoving frame.
The next step (in principle) is to apply the $L_{\rm DSR}$ to obtain the general form of EMT in DSR.
We f\/irst def\/ine the nonlinear mapping for the energy-momentum tensor of a~perfect f\/luid in a~comoving frame.
In the second step we shall apply the Lorentz boost ($L_{\rm SR}$) on our mapped variable and f\/inally arrive at the desired
expression in the DSR spacetime through an inverse mapping.
But we will see that when we try to introduce the f\/luid variables in the DSR EMT in arbitrary frame we face
a~non-trivial problem unless we make some simplifying assumptions, which, however, will still introduce DSR corrections
pertaining to the Planck scale cutof\/f.

As the spherical symmetry remains intact in the DSR theory~\cite{sg} we def\/ine the respective components of
energy-momentum tensor $\tilde{t}^{\mu \nu}$ in the NC framework analogous to~\eqref{lD}, \eqref{lN}
as
\begin{gather}
\tilde{t}^{ii} = p = \frac{1}{3} \sum\limits_{i}\frac{{\bf p}_{i}^{2}}{p^0_{i}}\delta^{3} (x-x_i),
\qquad
\tilde{t}^{00} = \rho = \sum\limits_{i} p^0_i\delta^{3} (x - x_i ),
\qquad
n = \sum\limits_{i}\delta^{3} (x - x_i),\!\!\!
\label{lr1}
\end{gather}
where ${\bf p}_i$ and $p^0_i$ are respectively the momentum three-vector and the energy of the~$i$-th f\/luid particle in
the DSR spacetime.
Using~\eqref{lr1} and using the scaling properties of Dirac-$\delta$ function we obtain the following results
\begin{gather}
F^{-1}(p) = \frac{1}{3} \sum\limits_{i}\frac{{\bf P}_{i}^{2}}{P^0_{i}\big(1 + P^0_i/\kappa\big)^4}\delta^{3} (X -
X_i),
\label{le}
\\
F^{-1}(\rho) = \sum\limits_{i} \frac{P^0_i}{\big(1 + P^0_i/\kappa\big)^4}\delta^{3} (X - X_i ),
\label{lf}
\\
F^{-1}(n) = \sum\limits_{i} \frac{N}{\big(1 + P^0_i/\kappa \big)^3}\delta^{3} (X - X_i ).
\label{ln}
\end{gather}
In a~combined form, we can write down the following nonlinear mapping (and its inverse) as
\begin{gather}
F^{-1}\big(\tilde{t}^{\mu\nu}\big) = \sum\limits_{i}\frac{P^\mu_{i} P^\nu_{i}}{P^0_{i}\big(1 + P^0_i/\kappa\big)^4}\delta^{3} (X-X_i),
\nonumber
\\
F\big(\tilde{T}^{\mu\nu}\big) = \sum\limits_{i}\frac{p^\mu_{i}p^\nu_{i}}{p^0_{i}\big(1 + p^0_i/\kappa\big)^4}\delta^{3} (x-x_i).
\label{ft}
\end{gather}
The way we have def\/ined the DSR EMT it is clear that comoving form of EMT also receives DSR corrections.
But problem crops up when, in analogy to SR EMT~\cite{wein}, we attempt to boost the $\tilde t^{\mu\nu}$ to a~laboratory
frame with an arbitrary velocity~$v^i$.
Recall that for a~single particle DSR boosts involve its energy and momentum.
Since~$p$ and~$\rho$ (for $\tilde t^{\mu\nu}$) denote composite variables it is not clear which energy or momentum will
come into play.
To proceed further, in the expression of DSR boost, we put in a~single energy $\bar p^0$ and momentum $\bar p^i$ which
denotes the average energy and momentum (modulus) of the whole f\/luid.
In fact this simplif\/ication is not very artif\/icial since we are obviously considering equilibrium systems in our study.
This allows us to use the mappings
\begin{gather*}
F^{-1}(p) = \frac{P}{\big(1+\bar P^0/\kappa\big)^4},
\qquad
F^{-1}(\rho) = \frac{D}{\big(1+\bar P^0/\kappa\big)^4},
\qquad
F^{-1}(n) = \frac{N}{\big(1+\bar P^0/\kappa\big)^4}.
\end{gather*}
In a~covariant form the mapping and its inverse between $\tilde{t}^{\mu \nu}$ and $\tilde{T}^{\mu \nu}$ are
\begin{gather*}
F^{-1}\big(\tilde{t}^{\mu \nu}\big) = \frac{\tilde{T}^{\mu \nu}}{\big(1+\bar P^0/\kappa\big)^4},
\qquad
F\big(\tilde{T}^{\mu \nu}\big) = \frac{\tilde{t}^{\mu \nu}}{\big(1-\bar p^0/\kappa\big)^4}.
\end{gather*}
Finally we can apply the def\/inition of $L_{\rm DSR}$ to obtain the following expressions for energy-momentum tensor with
respect to an arbitrary inertial frame
\begin{gather*}
t^{00} = L_{\rm DSR}\big(\tilde{t}^{00}\big) = F\circ L_{\rm SR}\circ F^{-1}\big(\tilde{t}^{00}\big) = F \circ L_{\rm SR}\left(\frac{\tilde{T}^{00}}{\big(1+\bar P^0/ \kappa\big)^4}\right)
\\
\phantom{t^{00}}
= F \left(\frac{\gamma^2\big(D+P v^2\big)}{\big(1+\frac{\gamma}{\kappa}\big(\bar P^0-v\bar P^1\big)\big)^4}\right) =
\frac{\gamma^2\big(\rho+p v^2\big)}{\bar \alpha^4},
\\
t^{i0} = L_{\rm DSR}\big(\tilde{t}^{i0}\big) = \frac{\gamma^2(\rho+p)v^i}{\bar \alpha^4},
\qquad
t^{ij}=L_{\rm DSR}\big(\tilde{t}^{ij}\big) = \frac{\gamma^2(\rho+p)v^i v^j}{\bar \alpha^4}+p \delta^{ij}.
\end{gather*}
It is very interesting to note that the above expressions can also be combined into a~single form which is structurally
very close to the f\/luid EMT in SR
\begin{gather}
t^{\mu\nu}=\frac{\left(1-\frac{\bar p_0}{\kappa}\right)^2}{\bar \alpha^{4}} \left((p+\rho)u^{\mu}u^{\nu}+
p\frac{\eta^{\mu\nu}}{\big(1-\frac{\bar p_0}{\kappa}\big)^2}\right)= \frac{\big(1-\frac{\bar
p_0}{\kappa}\big)^2}{\bar \alpha^{4}} \big((p+\rho)u^{\mu}u^{\nu}+ p\tilde{\eta}^{\mu\nu}\big),
\label{lnctf}
\end{gather}
where we have def\/ined the four-velocity $u^\mu$ in the DSR spacetime as
\begin{gather*}
u^0 = dx^0/d\tau = \frac{\gamma}{\left(1-\bar p_0/\kappa\right)},
\qquad
u^i = dx^i/d\tau = \frac{\gamma v^i}{\left(1-\bar p_0/\kappa\right)}.
\end{gather*}
Note that the DSR four-velocity $u^\mu$ is actually the mapped form of the SR four-velocity~$U^\mu$ since the parameter~$\tau $ does not undergo any transformation.
The other point to notice is that~$\tilde\eta^{\mu\nu}$ of~\eqref{ncm}, (DSR analogue of the f\/lat metric~$\tilde\eta^{\mu\nu}$), appears in $t^{\mu\nu}$ making the f\/inal form of the DSR EMT transparent.
Indeed $t^{\mu\nu}$ in~\eqref{lnctf} reduces smoothly to $T^{\mu\nu}$ of SR~\eqref{lctf} in the large~$\kappa$ limit.
Incidentally, again in analogy to the SR construction of many-body system for f\/luid (\eqref{lt1},~\eqref{lD})
this form of $t^{\mu\nu}$ is consistent with the microscopic picture of DSR EMT for f\/luid that we have developed
(\eqref{lr1}--\eqref{ft}).
Derivation of this DSR-covariant expression of energy-momentum tensor~\eqref{lnctf} is the major result of our
work~\cite{dasdsr}.

\section[Thermodynamics of photon gas in~$\kappa$-Minkowski spacetime]{Thermodynamics of photon gas in~$\boldsymbol{\kappa}$-Minkowski spacetime}
\label{Section4}

We consider here a~particular modif\/ied dispersion relation in DSR, the Magueijo--Smolin (MS) dispersion
relation~\cite{dasdsr, dasph, leejoao, ms}
\begin{gather}
\epsilon^2 - p^2 = m^2 \left(1 - \frac{\epsilon}{\kappa}\right)^2,
\label{ms1}
\end{gather}
where $p_{\mu} = (\epsilon, \vec{p})$ is the DSR four-momentum and $p \equiv |\vec{p}|$ is the magnitude of the
three-momentum of a~particle.
Thermodynamic properties for photon gas with a~dif\/ferent dispersion relation have been studied in~\cite{camacho1}.
Also, thermodynamics of bosons and fermions with another modif\/ied dispersion relation and its cosmological and
astrophysical implications have been observed in~\cite{bertolami, magcos}.
But these two modif\/ied dispersion relations appear from a~phenomenological point of view whereas the dispersion
relation~\eqref{ms1} has a~more theoretical motivation which we discuss below in some details.

It was shown in~\cite{sg} that existence of an invariant length scale in the theory is consistent with a~noncommutative
(NC) phase space ($\kappa$-Minkowski spacetime) such that the usual cano\-ni\-cal Poisson brackets between the phase space
variables are modif\/ied.
Also, the linear Lorentz transformations in special relativity ($L_{\rm SR}$) are replaced by nonlinear DSR-Lorentz
transformations ($L_{\rm DSR}$)~\cite{br, sg}.
One point about our notation convention: throughout the rest of this Section, small letters like $x$, $p$ denote DSR
variables whereas capital letters $X$, $P$ denote the corresponding variables in SR theory.
Now, using this NC phase space algebra in DSR, one can readily check that the Lorentz algebra is intact
\begin{gather*}
\big\{j^{\mu\nu},j^{\alpha\beta}\big\} = g^{\mu\beta} j^{\nu\alpha} +g^{\mu\alpha}j^{\beta \nu} + g^{\nu\beta}j^{\alpha\mu}
+g^{\nu\alpha}j^{\mu\beta},
\end{gather*}
where the angular momentum is def\/ined in the usual way as
\begin{gather*}
j_{\mu\nu} = x_\mu p_\nu - x_\nu p_\mu.
\end{gather*}
As a~result, we have the $L_{\rm DSR}$ invariant modif\/ied dispersion relation~\eqref{ms1} as
\begin{gather*}
\left\{j_{\mu \nu},
\frac{p^2}{\left(1-\frac{\epsilon}{\kappa} \right)^2}\right\} = 0.
\end{gather*}
Due to the nontrivial expression for the dispersion relation~\eqref{ms1}, f\/irstly it was supposed that the velocity of
photon $c=\frac{d\epsilon}{d p}$
have to be
energy dependent.
But it was shown in~\cite{hoss} that a~modif\/ied dispersion relation does not necessarily imply a~varying (energy
dependent) velocity of light.
Thus, though the above two models (\cite{camacho1} and~\cite{bertolami, magcos}) admit a~varying speed of light, in case of the Magueijo--Smolin (MS) DSR model considered
here, for photons ($m=0$) the dispersion relation~\eqref{ms1} is the same as in SR theory.
Also the speed of light~$c$ is an invariant quantity in the DSR model~\cite{dasdsr, sg, visser, leejoao, ms}.
Thus the DSR model considered in~\cite{sg, leejoao, ms} has a~more theoretical motivation and it can be developed
starting from the NC phase space variables~\cite{sg} whereas the models considered in~\cite{bertolami, camacho1, magcos}
are phenomenological in nature and as far as we know, there is no fundamental phase space structures to describe these models.

Another interesting fact is that both the models described in~\cite{camacho1} and in~\cite{bertolami, magcos} have no
f\/inite upper bound of energy of the photons though they have a~momentum upper bound.
But, as stated earlier, in the Magueijo--Smolin (MS) case, though the dispersion relation for the photons is unchanged,
there is a~f\/inite upper bound of photon energy which is the Planck energy~$\kappa$.
One can readily check that this is an invariant quantity by using the DSR-Lorentz transformation law for
energy~\cite{br, sg}.

One more thing must be clarif\/ied here.
In case of the models (\cite{camacho1} and~\cite{bertolami, magcos}), clearly the Lorentz symmetry was broken and as
a~result, the number of microstates and hence the entropy increases as compared to the Lorentz symmetric SR theory.
On the other hand, we are dealing with a~dif\/ferent scenario where the Lorentz symmetry is not broken as Lorentz algebra
between the phase space variables is intact.
In fact, the framework we describe here still satisf\/ies the basic postulates of Einstein's SR theory; moreover it
possesses another observer independent quantity.
Thus it seems that Lorentz symmetry is further restricted in this DSR model.
As a~result of this, we expect to have a~less number of microstates and less entropy in the MS model.
As we will show later in our explicit calculations, this expected result is correct.

\subsection{Partition function for photon gas}

To study the thermodynamic behavior of photon gas, we have to f\/ind out an expression for the partition function f\/irst,
as it relates the microscopic properties with the thermodynamic (macroscopic) behavior of a~physical
system~\cite{greiner, pathria}, which we do in this section.
As we have said earlier, the modif\/ied dispersion relation~\eqref{ms1} in case of the photons (massless particles) does
not change from the usual SR scenario.
Thus, for the photons, the dispersion relation now becomes
\begin{gather*}
|\vec{p}| = p = \epsilon.
\end{gather*}
We consider a~box containing photon gas.
Following the standard procedure as given in~\cite{greiner, pathria}, we consider a~continuous spectrum of momentum
instead of quantizing it.
The number of microstates available to the system ($\sum$) in the position range from~$r$ to $r+dr$ and in the momentum
range from~$p$ to $p+dp$ is given~by
\begin{gather*}
\sum = \frac{1}{h^3} \iint d^3 \vec{r} d^3 \vec{p},
\end{gather*}
where~$h$ is the phase space volume of a~single lattice and
\begin{gather*}
\iint d^3 \vec{r} d^3 \vec{p}
\end{gather*}
is the total phase space volume available to the system.

It should be mentioned here that as in the case of SR theory~\cite{misner}, the phase space volume element $d^3 x d^3 p$
in DSR is also a~DSR-Lorentz invariant quantity (for details, please see~\cite{dasph}).
If the volume of the box is considered to be~$V$, in case of SR, the number of microstates can be written in the
following form using the spherical polar coordinates
\begin{gather}
\sum = \frac{4 \pi V}{h^3} \int_0^{\infty} E^2 d E.
\label{state2}
\end{gather}
We used the dispersion relation $P=E$ to change the integration variable to~$E$.
Now, in case of the DSR model, considering the fact that we have an f\/inite upper limit of energy ($\kappa$), we obtain
the number of microstates as
\begin{gather}
{\tilde{\sum}} = \frac{4 \pi V}{h^3} \int_0^{\kappa} \epsilon^2 d \epsilon,
\label{state}
\end{gather}
where $\tilde{\sum}$ represents the number of microstates in the DSR model which we have considered here.
It is obvious from the expressions~\eqref{state2} and~\eqref{state} that the available number of microstates to the
system in case of DSR is less than that in the SR theory.
This happens due to the fact that there is an upper energy bound in the DSR model whereas the energy spectrum of
a~particle in SR theory can go all the way up till inf\/inity.
This result agrees with our expectation as stated earlier.

It is very crucial to get an expression for the partition function as all the thermodynamic properties can be thoroughly
studied using the knowledge about the partition function.
For the conventional free particle in SR, the partition function $Z_1(T, V)$ is def\/ined as~\cite{greiner}
\begin{gather}
Z_1(T, V) = \frac{4 \pi V}{h^3} \int_0^{\infty} P^2 e^{-\frac{E}{k_{\rm B} T}} d P,
\label{partition1}
\end{gather}
where $k_{\rm B}$ is the Boltzman constant and~$T$ is the temperature.

For our DSR model, the single particle partition function $\tilde{Z}_1(T, V)$ is def\/ined as
\begin{gather}
\tilde{Z}_1(T, V) = \frac{4 \pi V}{h^3} \int_0^{\kappa} p^2 e^{-\frac{\epsilon}{k_{\rm B} T}} d p.
\label{partition1dsr}
\end{gather}
In the limit $\kappa \rightarrow \infty$, we get back normal SR theory results.

It should be noted that in the DSR model which we have considered, the photon dispersion relation is not modif\/ied at all
(as for massless photons, $p=\epsilon$).
But still there is modif\/ication in the partition function~\eqref{partition1dsr} due to the presence of an energy upper
bound of particles ($\kappa$) in the theory.
Note that the upper limit of integration is~$\kappa$ in~\eqref{partition1dsr} whereas in the normal SR theory
expression~\eqref{partition1}, the upper limit of integration is inf\/inity.
In all the models~\cite{bertolami, camacho1, magcos}, though the upper limit of energy is inf\/inity as in SR theory,
these models have dif\/ferent dispersion relations than~SR.

Using the dispersion relation for photons ($\epsilon = p$) and using the standard table and formulae for
integrals~\cite{gradstein}, we f\/inally have an analytic expression of the single particle partition function
\begin{gather}
\tilde{Z}_1(T, V) = \frac{4 \pi V}{h^3} \int_0^{\kappa} \epsilon^2 e^{-\frac{\epsilon}{k_{\rm B} T}} d \epsilon
\nonumber
\\
\phantom{\tilde{Z}_1(T, V)}
= \frac{4 \pi V}{h^3} \left[2 (k_{\rm B} T)^3 - e^{-\frac{\kappa}{k_{\rm B} T}} (k_{\rm B} T)^3
\left(2 + \frac{\kappa}{k_{\rm B} T} \left(2+\frac{\kappa}{k_{\rm B} T}\right)\right)\right].
\label{part1}
\end{gather}
Thus the partition function for a~$N$-particle system $\tilde{Z}_N(T, V)$ is given~by
\begin{gather}
\tilde{Z}_N(T, V) = \frac{1}{N!} \big[\tilde{Z}_1(T, V)\big]^N,
\label{partn}
\end{gather}
where we have considered classical (Maxwell--Boltzman) statistics along with the Gibb's factor.

Describing a~multi-particle system in a~relativistically invariant way is a~non-trivial issue and more so in case of DSR
framework, where the momenta do not add up linearly.
Probably the best setup to discuss these issues is the relative locality~\cite{relocal,relocal1, maghoss, hoss1,
hossrev} framework.
Being not so ambitious we provide a~more simple prescription of essentially following the normal statistical mechanics
approach used for a~system of of non-interacting particles.
This means that the partition function of the multi-particle system is that of the single particle system raised to the
power of~$N$, the number of particles.
The justif\/ication of our scheme is the following.
First of all note that, (also advocated in relative locality perspective~\cite{relocal,relocal1}) our system consists
of individual ``elementary particles'' (in the classical sense) for which normal special theory rules should apply.
Secondly whatever DSR corrections are considered they concern individual particle momenta and the interaction terms are
damped by a~factor of $N M_{\rm P}$ ($M_{\rm P}$ being the Planck mass), which is macroscopic for a~thermodynamic system.
Furthermore in~\cite{relocal1} the need for an appropriate coordinate system has been emphasized.
Clearly one such coordinate system is the canonical (Darboux) coordinates, provided in~\cite{sg}.
Expressing the partition function in the canonical coordinates and then reverting back to the physical coordinates we
can argue that the DSR ef\/fects manifest only in single particle partition function which is characterized here by the
upper limit of the energy,~$\kappa$ in the energy integral, a~signature of DSR models.
For these reasons we expect that partition function constructed here for a~DSR photon gas will hold to lowest order of~$\kappa$.

\subsection{Thermodynamic properties of photon gas}

With the expression for the partition function~\eqref{part1}, \eqref{partn} in hand, now one can study various
thermodynamic properties of the photon gas for this DSR model.
It should be noted that as $\kappa \rightarrow \infty$, this partition function coincides with the partition function in
SR theory and thus all of our results coincides with the usual SR case in this limit.

We use Stirling's approximation for $ \ln [N!]$~\cite{greiner}
\begin{gather*}
\ln [N!] \approx N \ln[N] - N
\end{gather*}
in the expression for partition function~\eqref{partn} to obtain the free energy $\tilde{F}$ of the system
\begin{gather}
\tilde{F} = -k_{\rm B} T \ln\big[\tilde{Z}_N(T, V)\big]
\nonumber
\\
\phantom{\tilde{F}}
= - N k_{\rm B} T \left[1 + \ln \left[\frac{4 \pi V}{N}\left(\frac{k_{\rm B} T}{h}\right)^3 \left\{2 - e^{- \frac{\kappa}{k_{\rm B}
T}}\left(2 + \frac{\kappa}{k_{\rm B} T}\left(2 + \frac{\kappa}{k_{\rm B} T}\right)\right)\right\}\right] \right].
\label{free1}
\end{gather}
In the limit $\kappa \rightarrow \infty$, the terms containing~$\kappa$ vanishes and we get back normal SR theory
result: $F=-N k_{\rm B} T$.

From the expression for free energy~\eqref{free1} we can readily obtain the expression for entropy $\tilde{S}$ of photon
gas in our considered DSR model as~\cite{greiner}
\begin{gather}
\tilde{S} = -\left(\frac{\partial \tilde{F}}{\partial T}\right)_{V, N}
= N k_{\rm B} \left[4 + \ln \left[\frac{4 \pi V}{N}\left(\frac{k_{\rm B} T}{h}\right)^3 \left\{2 - e^{- \frac{\kappa}{k_{\rm B} T}}\left(2
+ \frac{\kappa}{k_{\rm B} T}\left(2 + \frac{\kappa}{k_{\rm B} T}\right) \right) \right\}\right] \right.
\nonumber
\\
\phantom{\tilde{S}=}
\left.
- \frac{\kappa^3}{2 k_{\rm B}^3 T^3 e^{\frac{\kappa}{k_{\rm B} T}} - \big(2 k_{\rm B}^3 T^3 + 2 k_{\rm B}^2 T^2 \kappa + k_{\rm B} T
\kappa^2\big)}\right].
\label{entropy}
\end{gather}
The terms containing~$\kappa$ in the above expression~\eqref{entropy} are the DSR modif\/ication terms.
In the limit $\kappa \rightarrow \infty$ the terms containing~$\kappa$ vanish and we get back the SR theory result
\begin{gather*}
S = N k_{\rm B} \left[4 + \ln \left[\frac{8 \pi V}{N}\left(\frac{k_{\rm B} T}{h}\right)^3 \right] \right].
\end{gather*}
We plot the entropy~$S$ against temperature~$T$ both for the DSR model and for SR theory to study the deviation of
entropy in the two models.

\begin{figure}[t]
\centering \includegraphics[width = 11cm]{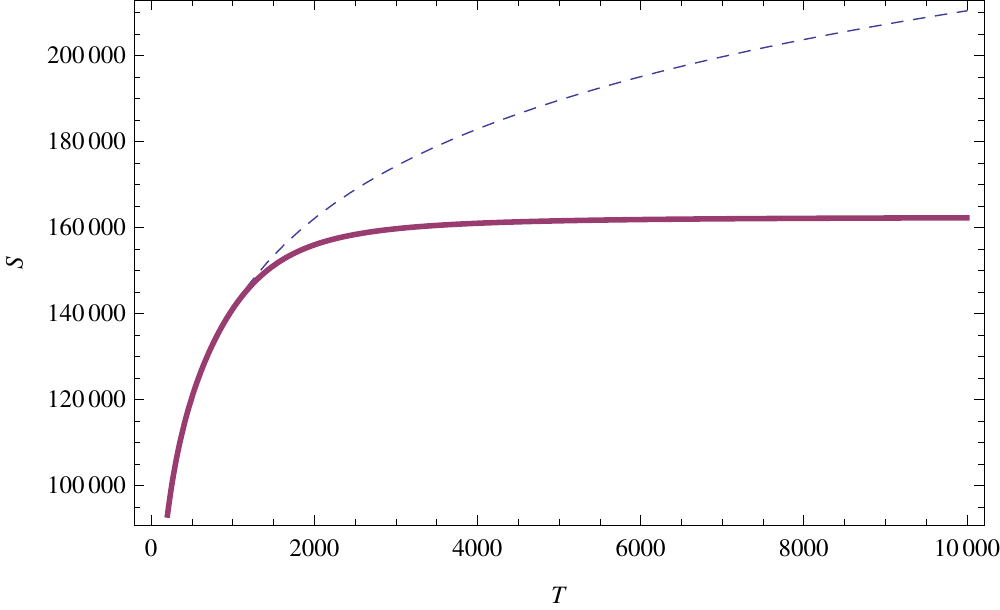}
\caption{Plot of entropy of photon~$S$ against
temperature~$T$ for both in the SR theory and DSR model; the dashed line corresponds to the SR theory result and the
thick line represents the corresponding quantity in DSR.
We have used the Planck units and the corresponding parameters take the following values $\kappa = 10000$, $k_{\rm B} = 1$,
$N = 10000$, $V =0.01$, $h = 1 $ in this plot.
With this scale, $T=10000$ represents the Planck temperature.}
\label{Fig1}
\end{figure}

In Fig.~\ref{Fig1}, we have plotted entropy against temperature for both the case of DSR and usual SR theory.
It is clearly observable from the plot that the entropy grows at a~much slower rate in case of DSR than in the SR theory
and as temperature increases, the entropy in DSR model deviates more from the entropy in the SR theory.
This result matches with our earlier expectation considering the underlying symmetry of the theory that the entropy in
the DSR model should be less than the entropy in SR theory.
As $T=10000$ is the Planck temperature, from the above plot one can see that the entropy saturates well before reaching
the Planck scale (nearly around $T=2000$).
However, this saturation temperature is still very much high to experimentally observe these DSR ef\/fects.

It is well known that the total number of microstates available to a~system is a~direct measure of the entropy for that
system.
Therefore our result merely ref\/lects the fact that due to the existence of an energy upper bound~$\kappa$ in the DSR
model, the number of microstates gradually saturates to some f\/inite value.

We expect modif\/ication in the expression of the internal energy~$U$ for photon gas in the DSR model as the expression of
entropy is modif\/ied and internal energy is related to the entropy as follows: $U = F + T S$.
In the usual SR scenario, the explicit expression for internal energy is given by $U = 3 N k_{\rm B} T$. But in the DSR
scenario we considered, the expression for internal energy ($\tilde{U}$) of photon gas is the following
\begin{gather}
\tilde{U} = N k_{\rm B} T \left[3 - \frac{\kappa^3 e^{- \frac{\kappa}{k_{\rm B} T}}}{2 k_{\rm B}^3 T^3 - e^{- \frac{\kappa}{k_{\rm B} T}}\big(2
k_{\rm B}^3 T^3 + 2 \kappa k_{\rm B}^2 T^2 + \kappa^2 k_{\rm B} T\big)}\right].
\label{internaldsr}
\end{gather}
It is easy to see from the expression of internal energy~\eqref{internaldsr} that we get back the usual SR theory
expression in the limit $\kappa \rightarrow \infty$.
As in the case of entropy, here also we plot internal energy against temperature for both the SR and DSR case.

\begin{figure}[t]
\centering \includegraphics[width = 11cm]{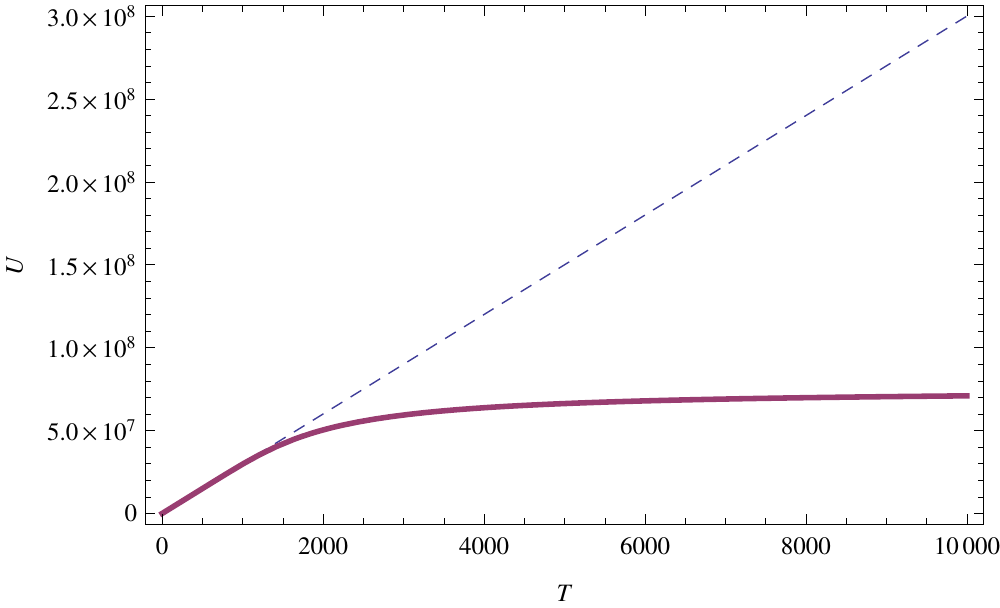}
\caption{Plot of internal energy of photon against
temperature for both in the SR theory and DSR scenario; the dashed line corresponds to the SR theory result and the
thick line represents the quantity in the DSR model.
We used the Planck units and the corresponding parameters take the following values
$\kappa = 10000$, $k_{\rm B} = 1$, $N=10000$, $V =0.01$,
$h = 1$ in this plot.
With this scale, $T=10000$ represents the Planck temperature.}
\label{Fig2}
\end{figure}

In Fig.~\ref{Fig2}, we plotted internal energy of photon gas against its temperature for both the case of DSR model and
SR theory.
One can easily see from the plot that the value of internal energy (for a~particular temperature) in the DSR
model~\eqref{internaldsr} is always less than its value (for the same temperature) in the SR theory.
Since the internal energy~$U$ of photon gas becomes saturated after a~certain temperature in case of the DSR model, it
is tempting to point out that probably our results are moving towards the right direction related to the ``soccer ball
problem'' that plagues multi-particle description in the framework of DSR~\cite{maghoss, hossrev}.

In the next two sections, we discuss GUP ef\/fects for dif\/ferent types of scenarios.

\section{Covariant formulation of GUP Lagrangian}\label{Section5}

{\sloppy Operatorial forms of noncommutative (NC) phase space structures has the generic form
\begin{gather}
\{x_i, p_j\}=\delta_{ij}\big(1 + f_1\big({\mathbf p}^2\big)\big) + f_2\big({\mathbf p}^2\big) p_i p_j,
\nonumber
\\
\{x_i, x_j\}=f_{ij}({\mathbf p}),
\qquad
\{p_i,p_j\}=g_{ij}({\mathbf p}),
\qquad
i=1,2,3.
\label{xp}
\end{gather}
Interestingly, potential application of~\eqref{xp} can generate generalized uncertainty principle (GUP) which is
compatible with string theory expectation~\cite{gup1, gup3, maggiore1, maggiore2, str} that there exist a~minimum length scale or a~maximum momentum in nature.
Such a~length scale is def\/ined to be of the order of ${\sqrt{\beta}}$, where $\beta $ can be treated as a~small parameter.
The corresponding models of GUP have been proposed in a~non-covariant framework, by Kempf~\cite{kem}
(a~two-parameter model, with~$\beta$ and $\beta'$)
\begin{gather}
\{x_i,p_j\}=\delta_{ij}\big(1 +\beta {\mathbf p}^2\big) +\beta' p_i p_j,
\nonumber
\\
\{x_i,x_j\}=(\beta'-2\beta)(x_ip_j-x_jp_i),
\qquad
\{p_i,p_j\}=0
\label{ppcommutator}
\end{gather}
by Kempf, Mangano and Mann~\cite{kem1}
\begin{gather}
\{x_i,p_j\}=\delta_{ij}\big(1 + \beta {\mathbf p}^2\big),
\qquad
\{x_i,x_j\}=-2\beta (x_ip_j-x_jp_i),
\qquad
\{p_i,p_j\}=0,
\label{ppcommutator1}
\end{gather}
and also by Kempf and Mangano~\cite{kem2}
\begin{gather}
\{x_i,p_j\}=\frac{\beta {\mathbf p}^2\delta_{ij}}{\sqrt{\big(1+2\beta {\mathbf p}^2\big)}-1}+\beta p_i p_j,
\qquad
\{x_i,x_j\}=0,
\qquad
\{p_i,p_j\}=0.
\label{ppcommutator2}
\end{gather}
The f\/irst NC algebra proposed by Snyder~\cite{sn} has the same structure that of~\eqref{ppcommutator1}.
In fact~\eqref{ppcommutator}~\cite{kem} and~\eqref{ppcommutator1}~\cite{kem1} can be reduced to the Snyder NC
form~\cite{sn} as discussed in~\cite{q}.
Here we restrict ourselves to the classical counterpart of the commutator algebra~\eqref{ppcommutator2}~\cite{kem2}
since it is structurally the simplest as the coordinates and momenta commute among themselves respectively.
But the results derived here are applied to quantum commutators as well.

}

We will consider a~relativistically covariant generalization of the algebra~\eqref{ppcommutator2}.
Starting with this algebra, f\/irstly we study a~generalized point particle Lagrangian~\cite{pram1} with a~non-canonical
symplectic structure that is equivalent to~\eqref{ppcommutator2}.
Latter on by introducing electrodynamic interaction term in Lagrangian we further study point particle
dynamics~\cite{pram1}.
Now from a~physical point of view this type of an intuitive particle picture is very useful and appealing since we can
see how it dif\/fers from the conventional relativistic point particle.
Also this particle model can act as a~precursor to f\/ield theories in such non-canonical space.
Similar point particle symplectic formalisms have been adopted in other forms of operatorial NC algebras, such
as~$\kappa$-Minkowski algebra~\cite{banerjee1, banerjee2, others, sghosh3, sghosh2, sg, girelli, mignemi, pinzul},
relevant in doubly special relativity framework~\cite{plb510, doub, nv, rev} or very special relativity
algebra~\cite{dasvsr, gibbons}, proposed in~\cite{cohen}.
However, the crucial thing for one is to realize that the Jacobi identity is maintained by the linearized algebra~\cite{q}
\begin{gather*}
\{X_\mu,P_\nu\}= \delta_{\mu\nu}\big(1+\beta P^2\big)+2 \beta P_\mu P_\nu,
\qquad
\{P_\mu,P_\nu\} = \{X_\mu,X_\nu\} = 0
%\label{PPXXcom}
\end{gather*}
only to {$\cal{O}(\beta)$}.
If we consider $J(X_\mu,X_\nu,P_\lambda)$ to be of the operatorial form
\begin{gather*}
J(X_\mu, X_\nu, P_\lambda) = \{X_\mu,\{X_\nu,P_\lambda\}\}+\{P_\lambda,\{X_\mu, X_\nu\}\} +
\{X_\nu,\{P_\lambda,X_\mu\}\},
\end{gather*}
then we get
\begin{gather*}
J(X_\mu, X_\nu, P_\lambda) = 4 \beta^2 P^2(\delta_{\nu\lambda}P_\mu-\delta_{\mu\lambda}P_\nu).
\end{gather*}
But exact validity of Jacobi identity is quite imperative for the phase space algebra.
Furthermore, due to this violation of Jacobi, there can not be any point particle interpretation of this NC symplectic
structure.
This is due to the fact that the NC structures appear as Dirac brackets which always preserve Jacobi
identity~\cite{dirac}.
Therefore we will also construct deformed Poincar\'{e} generators that generate proper translations and rotations of the
variables.

\subsection{Covariantized point particle Lagrangian}

We begin by positing covariantized form of the NC algebra proposed in~\cite{kem2} in $3+1$ dimensions, with a~Minkowski
metric $\eta_{\mu\nu}\equiv(1,-1,-1,-1)$
\begin{gather}
\{x_\mu,p_\nu\} = -\frac{\beta p^2g_{\mu\nu}}{\sqrt{(1+2\beta p^2)}-1}-\beta p_\mu p_\nu \equiv -\Lambda g_{\mu\nu}-\beta p_\mu p_\nu,
\nonumber
\\
\{x_\mu,x_\nu\}=0,
\qquad
\{p_\mu,p_\nu\} = 0,
\label{ppc}
\end{gather}
where $\Lambda =\frac{\beta p^2}{\sqrt{(1+2\beta p^2)}-1}$.
We would like to interpret the above relations~\eqref{ppc} as Dirac brackets derived from a~constrained symplectic
structure.
In some sense we are actually moving in the opposite direction of the conventional analysis where the computational
steps are
\begin{gather*}
{\rm Lagrangian}~\rightarrow~{\rm constraints}~\rightarrow~{\rm Dirac~brackets}
\end{gather*}
or equivalently
\begin{gather*}
{\rm symplectic~structure}~\rightarrow~{\rm symplectic~matrix}~\rightarrow~{\rm symplectic~brackets}.
\end{gather*}
Interestingly the Dirac brackets and symplectic brackets turn out to be the same.
In this case our path of analysis will be
\begin{gather*}
{\rm symplectic~brackets}~\rightarrow~{\rm symplectic~matrix}~\rightarrow~{\rm Lagrangian}.
\end{gather*}
Following this path, the symplectic matrix can be formed using~\eqref{ppc} as
\begin{gather*}
\Gamma_{ab}^{\mu\nu}= \left[
\begin{matrix}
0 & -(\Lambda g^{\mu\nu}+\beta p^\mu p^\nu)
\\
(\Lambda g^{\mu\nu}+\beta p^\mu p^\nu) & 0
\end{matrix}
 \right].
\end{gather*}
Inverse of this matrix provides commutators between the constraints as
\begin{gather}
\Gamma^{ab}_{\nu\lambda}= \left[
\begin{matrix}
0 & \left(\dfrac{g_{\nu\lambda}}{\Lambda}-\dfrac{\beta p_\nu p_\lambda}{\Lambda^2{\sqrt{1+2\beta p^2}}}\right)
\\
-\left(\dfrac{g_{\nu\lambda}}{\Lambda}-\dfrac{\beta p_\nu p_\lambda}{\Lambda^2{\sqrt{1+2\beta p^2}}}\right) & 0
\end{matrix}
 \right]\equiv\big\{\Phi^a_\nu,\Phi^a_\lambda\big\}.
\label{inv-mat}
\end{gather}
Indeed there is no unique way but from the constraint matrix one can make a~judicious choice of the constraints and
subsequently guess a~form of the Lagrangian.
We do not claim the Lagrangian derived in this way is unique, but one can easily check that the derived Lagrangian
yields the same Dirac brackets that one posited at the beginning.
Now it is convenient to work in the f\/irst order formalism where both $x_\mu$ and $p_\mu$ are treated as independent
variables with the conjugate momenta, $\pi_\mu^x=\frac{\partial L}{\partial\dot{x^\mu}}$,
$\pi_\mu^p=\frac{\partial L}{\partial\dot{p^\mu}}$ satisfying $\{x_\mu,\pi_\mu^x \}=-g_{\mu\nu}$, $\{p_\mu,\pi_\mu^p\}=-g_{\mu\nu}$.
We obtain from~\eqref{inv-mat} the following set of constraints
\begin{gather*}
\Phi_\mu^1=\pi_\mu^x \approx 0,
\qquad
\Phi_\mu^2=\pi_\mu^p+\frac{x_\mu}{\Lambda}-\frac{\beta(x p)p_\mu}{\Lambda^2\sqrt{1+2\beta p^2}} \approx 0.
\end{gather*}
From this constraint structure we can f\/inally write down the cherished form of the point particle Lagrangian in the
f\/irst order formalism of $(x,p)$ as
\begin{gather}
L=-\frac{(x\dot{p})}{\Lambda}+\frac{\beta(x p)(p\dot{p})}{\Lambda^2\sqrt{1+2\beta p^2}}+\lambda \big(f(p^2)-m^2\big),
\label{lag}
\end{gather}
where~$\lambda$ is a~Lagrange multiplier.
This construction of the particle model Lagrangian is one of our major results~\cite{pram1}.
We have included a~mass-shell condition $f(p^2)-m^2=0$ where $f(p^2)$ denotes an arbitrary function that needs to f\/ixed.
The Lorentz generators get modif\/ied to
\begin{gather*}
j_{\mu\nu}=\frac{1}{\Lambda}(x_\mu p_\nu-x_\nu p_\mu),
\end{gather*}
such that correct transformation of the degrees of freedom are reproduced
\begin{gather*}
\{j_{\mu\nu},p_\lambda\}=g_{\mu\lambda}p_\nu-g_{\nu\lambda}p_\mu,
\qquad
\{j_{\mu\nu},x_\lambda\}=g_{\mu\lambda}x_\nu-g_{\nu\lambda}x_\mu.
\end{gather*}
Interestingly this $j_{\mu\nu}$ obeys the correct Lorentz algebra
\begin{gather}
\{j_{\mu\nu},j_{\alpha\beta}\}=g_{\mu\alpha}j_{\nu\beta}-g_{\mu\beta}j_{\nu\alpha}-g_{\nu\beta}j_{\alpha\mu}
+g_{\nu\alpha}j_{\beta\mu}.
\label{lor}
\end{gather}
Now since $\{j_{\mu\nu},p^2\} = 0$, any function of $p^2$ is Lorentz invariant.
But keeping translation invariance in mind, a~more natural choice of $f(p^2)$ would be $f(p^2)\rightarrow
\frac{p^2}{\Lambda^2}$ leading to a~modif\/ied mass shell condition $\frac{p^2}{\Lambda^2} - m^2 = 0$.
However this can be actually simplif\/ied to $p^2=M^2$, $M=m/\big(1-\frac{\beta m^2}{2}\big)$.

\subsubsection{Approximations leading to other algebras}
As we have explained at the beginning, approximating the full NC algebra~\eqref{ppc} is not the proper way to derive an
ef\/fective ${\cal O}(\beta)$ corrected dynamical system since, in particular with operatorial NC algebras, there is
always a~drawback that Jacobi identities might be violated.
The correct way is to approximate the system at the level of the Lagrangian because then we are assured that the ${\cal
O}(\beta)$ corrected NC brackets will also satisfy the Jacobi identities.

\textbf{${\cal O}(\beta)$ results.}
To the f\/irst order approximation of~$\beta$, the function~$\Lambda$ becomes
$\Lambda=1+\frac{1}{2}\beta p^2 + {\cal O}(\beta^2)$, using which the equation~\eqref{lag} provides the Lagrangian
$L_{(1)}$ (without the mass-shell condition) as
\begin{gather*}
L_{(1)} = -(x\dot{p})\left(1-\frac{1}{2}\beta p^2\right) + \beta(x p)(p\dot{p}) + {\cal O}\big(\beta^2\big).
\end{gather*}
The Dirac brackets turn out to be
\begin{gather*}
\{x^\mu,p^\nu\} = -\left[\frac{g_{\mu\nu}}{\left(1-\frac{\beta p^2}{2}\right)}
+\frac{\beta p_\mu p_\nu}{\left(1-\frac{3\beta p^2}{2}\right)\left(1-\frac{\beta p^2}{2}\right)}\right],
\qquad
\{x^\mu,x^\nu\}=\{p^\mu,p^\nu\}=0.
\end{gather*}
Notice that the algebra is still structurally similar as the exact one and the Snyder form with non-zero
$\{x_\mu,x_\nu\}$ has not appeared.
This agrees with previous results that the Snyder form is present only in ${\cal O}(\beta^2)$ or when more than
one~$\beta$-like parameters are present~\cite{kem1, q}.
However, linearizing this algebra to ${\cal O}(\beta)$ is once again problematic as it clashes with the Jacobi identity.
We will see that the Snyder form is necessary in the linearized system in order to exactly satisfy the Jacobi identity.

The combination $x_\mu$, $\big(1-\frac{\beta p^2}{2}\big)p_\nu$ constitutes
a~canonical pair with $\big\{x_\mu, \big(1-\frac{\beta p^2}{2}\big)p_\nu\big\}=-g_{\mu\nu}$.
The operator $j_{\mu\nu} = \big(1-\frac{\beta p^2}{2}\big)(x_\mu p_\nu - x_\nu p_\mu)$ transforms $x_\mu$ and $p_\mu$ correctly
and satisf\/ies the correct Lorentz algebra~\eqref{lor}.

\textbf{${\cal O}(\beta^2)$ results.}
With $\Lambda \approx 1+\frac{\beta p^2}{2}-\big(\frac{\beta p^2}{2}\big)^2$,
the Lagrangian $L_{(2)}$ (without the mass-shell condition) becomes
\begin{gather*}
L_{(2)}=-(x\dot{p})\left(1-\frac{\beta p^2}{2}+\left(\frac{\beta p^2}{2}\right)^2\right)+\beta(x
p)(p\dot{p})\left(1-\frac{3\beta p^2}{2}\right).
\end{gather*}
The corresponding Dirac brackets are
\begin{gather*}
\{x_\mu,x_\nu\}=D(x_\mu p_\nu-x_\nu p_\mu),
\qquad
\{p_\mu,p_\nu\}=0,
\\
\{x_\mu,p_\nu\}=-\frac{g_{\mu\nu}}{\Big(1-\frac{\beta p^2}{2}+\big(\frac{\beta p^2}{2}\big)^2\Big)}-C p_\mu p_\nu,
\end{gather*}
where
\begin{gather*}
C=\frac{\beta\big(1-\frac{3\beta p^2}{2}\big)}
{\big(1-\frac{3\beta p^2}{2}+\frac{7\beta^2 p^4}{4}\big)\big(1-\frac{\beta p^2}{2}+\frac{\beta^2 p^4}{4}\big)},
\qquad
D=\frac{C\beta p^2}{2\big(1-\frac{3\beta p^2}{2}\big)}.
\end{gather*}
We notice that the Snyder form has been recovered once ${\cal O}(\beta^2)$ contributions are introduced.
This GUP based Snyder algebra connection constitutes the other major result of our work~\cite{pram1}.
It is possible to construct the deformed Poincar\'{e} generators but the expressions are quite involved and not very
illuminating.

\textbf{Two parameter $(\beta,\beta')$ results.}
We now provide a~considerably simpler Lagrangian with two
parameters~$\beta$ and $\beta'$ that can induce the Snyder algebra.
Note that {\it{ab initio}} it would have been hard to guess this result as well as the explicit expressions for the
algebra but in our constraint framework this is quite straightforward.
From the constraint analysis that generates the Dirac brackets it is clear that we need a~non-vanishing $\{\phi_2^\mu,
\phi_2^\nu \}$ to reproduce a~non-vanishing $\{x^\mu, x^\nu \}$ bracket.
Thus the two non-canonical terms in $L_{(1)}$ must have dif\/ferent~$\beta$-factors to produce the desired ef\/fect.
Hence we consider the Lagrangian $L_{(\beta,\beta')}$ (without the mass-shell condition)
\begin{gather*}
L_{(\beta,\beta')} = -(x\dot{p})\left(1-\frac{\beta p^2}{2}\right) + \beta'(x p)(p\dot{p}).
\end{gather*}
The Dirac brackets are obtained as
\begin{gather*}
\{x_\mu,x_\nu\} = D\frac{(\beta-\beta')}{\beta'}(x_\mu p_\nu-x_\nu p_\mu),
\qquad
\{p_\mu,p_\nu\}=0,
\\
\{x_\mu,p_\nu\}=-\frac{g_{\mu\nu}}{\left(1-\frac{\beta p^2}{2}\right)}-D p_\mu p_\nu,
\end{gather*}
where $D=\frac{\beta'}{\big(1-\frac{\beta p^2}{2}-\beta' p^2\big)\big(1-\frac{\beta p^2}{2}\big)}$.
Clearly for $\beta=\beta'\rightarrow \{x_\mu,x_\nu\}=0$, thus leaving a~GUP like algebra.
We have not shown the deformed Poincar\'{e} generators which are quite complicated.

\subsection{Particle in external electromagnetic f\/ield}

We introduce minimally coupled $U(1)$ gauge interaction to the free GUP particle Lagran\-gian~\eqref{lag}
\begin{gather*}
L = -\frac{(x\dot{p})}{\Lambda}+\frac{\beta(x p)(p\dot{p})}{\Lambda^2\sqrt{1+2\beta p^2}}+\lambda\big(f\big(p^2\big)-m^2\big)+e(A\dot{x}).
\end{gather*}
Since the symplectic structure changes here, so we need to compute the new Dirac algebra though the procedure remains
the same.
Therefore in this case the Dirac brackets modif\/ied by the $U(1)$ interaction are given by
\begin{gather}
\{x_\alpha,x_\gamma\}^*=0,
\qquad
\{x_\alpha,p_\gamma\}^*=-(\Lambda g_{\alpha\gamma}+\beta p_\alpha p_\gamma),
\nonumber
\\
\{p_\alpha,p_\gamma\}^*=-e\Lambda(\Lambda F_{\alpha\gamma}+\beta p_a(F_{\alpha a}p_\gamma-F_{\gamma
a}p_\alpha)).
\label{diracPP}
\end{gather}
It is convenient to consider the relativistic Hamiltonian of the form
\begin{gather}
H=\frac{p^2}{m}-\sqrt{p^2}.
\label{rel-ham}
\end{gather}
Using the Dirac brackets~\eqref{diracPP} and the Hamiltonian~\eqref{rel-ham}, the Hamilton's equations of motion are
obtained as,
\begin{gather}
\dot{x}_\alpha=\left\{x_\alpha,\frac{p_\gamma p_\gamma}{m}-\sqrt{p_\gamma p_\gamma}\right\}^*
=-\frac{1}{m}\Lambda^2\sqrt{1+2\beta p^2}p_\alpha,
\nonumber
\\
\dot{p}_\alpha=\left\{p_\alpha,\frac{p_\gamma p_\gamma}{m}-\sqrt{p_\gamma p_\gamma}\right\}^*
=-\frac{e}{m}\Lambda^2\sqrt{1+2\beta p^2}p_\gamma F_{\alpha\gamma}.
\label{dotp}
\end{gather}
Keeping only ${\cal O}(e)$ terms we can eliminate~$p$, to get the modif\/ied Newton's law
\begin{gather}
\ddot{x}_\alpha=-\frac{e}{m}\Lambda^2\sqrt{1+2\beta p^2}\dot{x}_\gamma
F_{\alpha\gamma}=-\frac{e}{m}\Lambda^2\sqrt{1+2\beta m^2}\dot{x}_\gamma F_{\alpha\gamma}.
\label{equmotion}
\end{gather}
It is important to note that the dynamics in~\eqref{dotp} and~\eqref{equmotion} is {\it{exact}} for the GUP
parameter~$\beta$ although it is to the f\/irst order of~$e$.
Hence the dynamics remains qualitatively unchanged with a~renormalization of the charge.
The ${\cal O}(\beta)$ equation of motion is given~by
\begin{gather*}
\ddot{x}_\alpha=-\frac{e}{m}\big(1+2\beta m^2\big)\dot{x}_\gamma F_{\alpha\gamma}.
\end{gather*}

\section{Path integral formulation for GUP Hamiltonian}\label{Section6}

The Heisenberg uncertainty principle (HUP) says that uncertainty in position decreases with increasing uncertainty in energy $(\Delta
x\sim\frac{\hbar}{\Delta p})$.
But HUP breaks down for energies close to Planck scale, at which point the Schwarzschild radius becomes comparable to
Compton wavelength.
Higher energies result in a~further increase of the Schwarzschild radius, inducing the following relation: $\Delta
x\approx l_{\rm P}^{2}\frac{\Delta p}{\hbar}$.
Consistent with the above, the following form of GUP has been proposed, postulated to hold in all scales~\cite{kem1}
\begin{gather}
\Delta x_{i}\Delta p_{i}\geq\frac{\hbar}{2}\big[1+\beta\big(\Delta p^{2}+\langle p\rangle^{2}\big)+2\beta\big(\Delta p_{i}^{2}+\langle p_{i}\rangle^{2}\big)\big],
\qquad
i=1,2,3,
\label{deltaxi}
\end{gather}
where $[\beta]=({\rm momentum})^{-2}$
and we will assume that $\beta \approx 1/(M_{\rm P} c)^{2} = l_{\rm P}^{2}/2\hbar^{2}$, $M_{\rm P}=$
Planck mass, and $M_{\rm P} c^{2}=$ Planck energy $\approx 10^{19}$~GeV.
In one dimension the above inequality takes the form
\begin{gather}
\Delta x\Delta p\geq\frac{\hbar}{2}\left[1+3\beta(\Delta p^{2}+\langle p\rangle^{2})\right]
\label{delta1}
\end{gather}
from which we have
\begin{gather}
\Delta p\leq \frac{\Delta x}{3\beta\hbar}+\sqrt{\left(\frac{\Delta x}{3\beta\hbar} \right)^{2}-\frac{1+3\beta\langle
p\rangle^{2}}{3\beta}}.
\label{deltafrac}
\end{gather}
Since $\Delta p$ is real quantity, we have
\begin{gather}
\left(\frac{\Delta x}{3\beta\hbar}\right)^{2}\geq\frac{1+3\beta\langle p\rangle^{2}}{3\beta} \ \Rightarrow \ \Delta x
\geq\hbar\sqrt{3\beta}\sqrt{1+3\beta\langle p\rangle^{2}},
\label{deltafrac2}
\end{gather}
which gives the minimum bound for $\Delta x$ as
\begin{gather}
\Delta x_{\min}=\hbar\sqrt{3\beta}.
\label{xmin}
\end{gather}
Here one should notice from~\eqref{deltafrac2} that the condition $\langle p\rangle = 0$ gives this minimum
bound~\eqref{xmin}, which is consistent with the above inequality~\eqref{deltafrac} and GUP relation~\eqref{delta1}.
Now using relation~\eqref{xmin} along with the condition $\langle p\rangle = 0$ we get the maximum bound of~$\Delta p$~as
\begin{gather}
\Delta p_{\max}=\frac{1}{\sqrt{3\beta}}.
\label{pmax}
\end{gather}
It can be shown~\cite{kem1} that the inequality~\eqref{deltaxi} follows from the modif\/ied Heisenberg algebra
\begin{gather}
[x_{i},p_{j}]=i\hbar\big(\delta_{ij}+\beta\delta_{ij}p^{2}+2\beta p_{i}p_{j}\big).
\label{heis}
\end{gather}
In this section, we use the notation $[\,,\,]$ instead of $\{\,,\,\}$ as used in the previous sections of this article
to denote brackets since we are now dealing with quantum Lie algebraic structure.
To satisfy the Jacobi identity, the above bracket~\eqref{heis} gives $[x_{i},x_{j}]=[p_{i},p_{j}]=0$, to f\/irst order in
${\cal O}(\beta)$~\cite{kem}.
Now def\/ining
\begin{gather}
x_{i} = x_{0i},
\qquad
p_{i} = p_{0i}\big(1+\beta p_{0}^{2}\big),
\label{x0i}
\end{gather}
where $p_{0}^{2}=\sum\limits_{j=1}^{3}p_{0j}p_{0j}$ and $x_{0i}$,~$p_{0j}$ satisfy
the canonical commutation relations
$[x_{0i},p_{0j}]=i\hbar\delta_{ij}$, one can easily show that the above commutation relation~\eqref{heis} is satisf\/ied,
to f\/irst order of~$\beta$.
Henceforth, we neglect terms of order $\beta^{2}$ and higher.
The ef\/fects of this GUP~\eqref{deltaxi} in Lamb shift and Landau levels have been studied in~\cite{gup5}.
Also, formulation of coherent states for this GUP has been described in~\cite{ghosh}.
Here we successfully derive the kernel for this GUP model by Hamiltonian path integral formulation~\cite{garrod} and
show that this GUP corrected kernel induces a~maximum momentum bound in the theory~\cite{pram2}.

Using~\eqref{x0i}, we start with the corresponding Hamiltonian of the form
\begin{gather*}
H=\frac{p^{2}}{2m} + V(\vec{r}),
\end{gather*}
which can be written as
\begin{gather}
H = H_{0} + H_{1} + {\cal O}\big(\beta^{2}\big),
\label{ham1}
\end{gather}
where
\begin{gather*}
H_{0} = \frac{p_{0}^{2}}{2m} + V (\vec{r}),
\qquad
H_{1} = \frac{\beta}{m}p_{0}^{4}.
\end{gather*}
Thus, we see that any system with an well def\/ined quantum (or even classical) Hamilto\-nian~$H_{0}$, is perturbed~by
$H_{1}$, near the Planck scale.
In other words, quantum gravity ef\/fects are in some sense universal.
Now the modif\/ied Schr\"odinger equation corresponding to the above Hamiltonian~\eqref{ham1} is given~by
\begin{gather*}
-\frac{\hbar^{2}}{2m}\frac{\partial^{2}}{\partial x^{2}} \psi(x,t)+\frac{\beta\hbar^{4}}{m}\frac{\partial^{4}} {\partial
x^{4}}\psi(x,t)+V(x)\psi=i\hbar\frac{\partial}{\partial t}\psi(x,t).
\end{gather*}
In this article we show that path integral method~\cite{feynman} is applicable to this higher energy cases and we
evaluate the free particle kernel for GUP corrected Hamiltonian~\eqref{ham1}.
For this purpose, we shall brief\/ly recall the notion of basic properties of kernel in Hamiltonian path integral
formalism.
The kernel in Hamiltonian path integral is given by~\cite{garrod}
\begin{gather}
K(x'', x')=\int\left[e^{\frac{i}{\hbar} \int(\vec{p}.
\vec{\dot{x}}-H)dt}\right]\frac{dp_{1}}{2\pi\hbar}
\frac{dp_{2}}{2\pi\hbar}
\cdots
\frac{dp_{N}}{2\pi\hbar}dx_{1}dx_{2}
\cdots
dx_{N-1}.
\label{kerhamdef}
\end{gather}
It has been shown in~\cite{garrod} that the above kernel~\eqref{kerhamdef} can be written in the form
\begin{gather*}
K(x'',x',\Delta t)=\delta (\vec{x}''-\vec{x}')-\frac{i\Delta t}{\hbar} \left[-\frac{\hbar^{2}\nabla^{2}}{2m}\times\delta
(\vec{x}''- \vec{x}')+\bar{V}(x)\delta (\vec{x}''-\vec{x}')\right]
\end{gather*}
from which one can easily obtain Schr\"odinger equation using the relation
\begin{gather}
\psi(x'',t'')=\int K(x'',t'';x',t')\psi(x',t')dx'.
\label{psi}
\end{gather}
Since
\begin{gather*}
\int \psi^*(x'',t'') \psi(x'',t'') d x'' = \int \psi^*(x',t') \psi(x',t') d x',
\end{gather*}
using the relation~\eqref{psi} we have
\begin{gather*}
\iiint  K^{*}(x'',t'';x'_{1},t')K(x'',t'';x',t')\psi^{*}(x'_{1},t') \psi(x',t')dx''dx'_{1} d x' =  \int
\psi^{*}(x',t') \psi(x', t') dx',
\end{gather*}
which immediately implies the following relation
\begin{gather}
\iint K^{*}(x'',t'';x'_{1},t')K(x'',t'';x',t')\psi^{*}(x'_{1},t')dx''dx'_{1}=\psi^{*}(x',t').
\label{kkf}
\end{gather}
Also, if $\psi(x,t)$ is the solution of the Schr\"odinger equation
\begin{gather*}
-\frac{\hbar^{2}}{2m}\nabla^{2}\psi(x,t)+V(x)\psi(x,t)=i\hbar\frac{\partial \psi(x,t)}{\partial t},
\end{gather*}
then the kernel also satisfy Schr\"odinger equation at the end point $x=x''$, i.e.
\begin{gather}
-\frac{\hbar^{2}}{2m}\frac{\partial^{2}}{{\partial x''}^{2}}K(x'',t'';x',t')+V(x'')K(x'',t'';x',t')=
i\hbar\frac{\partial}{\partial t''} K(x'',t'';x',t').
\label{schend}
\end{gather}
Equation~\eqref{psi},~\eqref{kkf} and~\eqref{schend} are the basic properties of the kernel $K(x,t)$.

\subsection{Kernel for GUP corrected Hamiltonian}
Path integral method~\cite{feynman} is applicable in all cases where the change of action, corresponding to the
variation of path, is large enough compared to $\hbar$.
As the above Hamiltonian~\eqref{ham1} is associated with higher energy, so a~small variation on paths other than the
path of least action make enormous change in phase for which cosine or sine will oscillate exceedingly rapidly between
plus and minus value and cancel out their total contribution.
So only the least action path will contribute in kernel.
This is similar to the idea of path integral in quantum mechanics.
We now therefore consider the Hamiltonian~\eqref{ham1} in one dimension
\begin{gather*}
H = \frac{p_{0}^{2}}{2m}+\frac{\beta}{m}p_{0}^{4}+V(x).
\end{gather*}
If we consider that the particle goes from $x'$ to $x''$ during the short time interval $\Delta t$, then the
inf\/initesimal kernel is of the form
\begin{gather}
K(x'',t'+\Delta t; x',t') = \int e^{\frac{i}{\hbar}\int_{t'}^{t'+\Delta t}\left(p_{0}.\dot{x}
-\frac{p_{0}^{2}}{2m}-\frac{\beta}{m}p_{0}^{4}-V(x)\right) d t}\frac{dp_{0}}{2\pi\hbar}
\nonumber
\\
\phantom{K(x'',t'+\Delta t; x',t')}
= \int e^{\frac{i}{\hbar}p_{0}.(x''-x')}e^{-\frac{i}{\hbar} \left[\frac{p_{0}^{2}\Delta t}{2m}+\frac{\beta\Delta
t}{m}p_{0}^{4}+\Delta t\bar{V}(x)\right]}\frac{dp_{0}}{2\pi\hbar},
\label{kerham}
\end{gather}
where $\bar{V}(x)$ is the average of $V(x)$ over the straight line connecting $x''$ and $x'$.
Expanding the second exponential function in~\eqref{kerham} and neglecting the second and higher order terms of $\Delta t$, we have
\begin{gather}
K(x'',t'+\Delta t; x',t') = \delta(x''-x')
\nonumber\\
\qquad{}
{}-\frac{i\Delta
t}{\hbar}\left[-\frac{\hbar^{2}}{2m}\frac{\partial^{2}}{{\partial
x''}^{2}}\delta(x''-x')+\frac{\beta\hbar^{4}}{m}\frac{\partial^{4}}{{\partial x''}^{4}}
\delta(x''-x')+\bar{V}(x)\delta(x''-x')\right].\label{kerham2}
\end{gather}
It is interesting to note that the kernel~\eqref{kerham2} boils down to the same form as in~\cite{garrod} in the limit $\beta \to 0$.
But it is very dif\/f\/icult to deal with above form of kernel~\eqref{kerham2} as it contains derivative of delta function.
Therefore we are going to derive the delta-independent equivalent form of kernel.
For this we consider the kernel for free particle
\begin{gather}
K(x'',t'+\Delta t; x',t')=\int e^{\frac{i}{\hbar}\int_{t'}^{t'+\Delta t}\left(p_{0}.\dot{x}
-\frac{p_{0}^{2}}{2m}-\frac{\beta}{m}p_{0}^{4}\right) d t}\frac{dp_{0}}{2\pi\hbar},
\label{kerfree}
\end{gather}
for short time interval $\Delta t$.
Now expanding the exponential series of the last term in~\eqref{kerfree} and neglecting the terms containing higher
order of~$\beta$ we have
\begin{gather*}
K(x'',t'+\Delta t; x',t')=\int e^{-\frac{i \Delta t}{2 m \hbar}\left(p_{0}-\frac{(x''-x')m}{\Delta t}\right)^{2}
+\frac{i m(x''-x')^{2}}{2\hbar\Delta t}}\left(1-\frac{i\beta\Delta t p_{0}^{4}}{\hbar m}\right)\frac{dp_{0}}{2\pi\hbar}.
\end{gather*}
After some calculation we get the kernel as
\begin{gather}
K(x'',t'+\Delta t; x',t')
\nonumber
\\
\qquad
=\sqrt{\frac{m}{2\pi i\hbar\Delta t}} \left[1+\frac{3\beta i\hbar m}{\Delta t} -\frac{6\beta
m^{2}(x''-x')^{2}}{{\Delta t}^{2}}-\frac{i\beta m^{3}(x''-x')^{4}} {\hbar{\Delta
t}^{3}}\right]e^{\frac{im(x''-x')^{2}}{2\hbar\Delta t}}.
\label{kerfreeinf}
\end{gather}
After a~bit of lengthy algebra (please see~\cite{pram2} for details), the f\/inal form of the kernel becomes
\begin{gather}
K(x'', t''; x', t')= \sqrt{\frac{m}{2 \pi i \hbar (t''-t')}}
\nonumber
\\
\hphantom{K(x'', t''; x', t')=}{}
\times e^{\frac{i m (x''-x')^2}{2 \hbar (t''-t')}}
\left[1+\frac{3\beta i\hbar m}{(t''-t')} -\frac{6\beta m^{2}(x''-x')^{2}}{(t''-t')^{2}}-\frac{i\beta m^{3}(x''-x')^{4}}
{\hbar(t''-t')^{3}}\right],\!\!\!\label{kerfreefinal}
\end{gather}
where $t''-t' = N \Delta t$.
This kernel~\eqref{kerfreefinal} is exactly of the same form as the kernel for the inf\/initesimal
interval~\eqref{kerfreeinf}.
It can be shown that the above kernel~\eqref{kerfreefinal} satisf\/ies the modif\/ied schrodinger equation
\begin{gather}
-\frac{\hbar^{2}}{2m}\frac{\partial^{2}}{\partial x^{2}}\psi(x,t) +\frac{\beta\hbar^{4}}{m}\frac{\partial^{4}}{\partial
x^{4}}\psi(x,t) =i\hbar\frac{\partial}{\partial t}\psi(x,t),
\label{kersch}
\end{gather}
at the point $x=x''$, $t=t''$.
Now the solution of this Schr\"odinger equation~\eqref{kersch} is given by~\cite{gup5}
\begin{gather}
\psi(x, t) = \Big(A e^{i k(1-\beta\hbar^{2}k^{2})x-\frac{i E t}{\hbar}}
\nonumber
\\
\phantom{\psi(x, t) =}
{}+ B e^{-i k(1-\beta\hbar^{2}k^{2})x-\frac{i E
t}{\hbar}} + C e^{\frac{x}{\sqrt{2\beta\hbar^{2}}}-\frac{i E t}{\hbar}} +De^{-\frac{x}{\sqrt{2\beta\hbar^{2}}}-\frac{i E
t}{\hbar}}\Big).
\label{psif}
\end{gather}
With this solution~\eqref{psif} in hand, we can show that the kernels~\eqref{kerham2} and~\eqref{kerfreeinf} indeed
pro\-pa\-gates the wave function $\psi(x, t)$ from a~point $(x', t')$ to the point $(x'', t'')$, for a~chosen time interval
$\Delta t = t''-t'$, such that $\frac{\Delta t}{4\beta\hbar m} = D$ = a~dimensionless quantity of ${\cal O}(\beta)$.
Thus the following relation holds
\begin{gather*}
\psi(x'', t'') = \int K (x'', t''; x',t') \psi(x', t') d x'.
\end{gather*}
Therefore the free particle kernel satisf\/ies the basic properties of a~kernel, which we have stated earlier.

For usual free particle case, the probability that a~particle arrives at the point $x''$ is proportional to the absolute
square of the kernel $K(x'',x',t''-t')$, i.e.~for usual free-particle kernel the probability is given~by
\begin{gather*}
P(x'') d x = \frac{m}{2 \pi \hbar (t''-t')} d x.
\end{gather*}
Now, for the GUP corrected kernel~\eqref{kerfreefinal}, the corresponding probability is given~by
\begin{gather}
P(x'') d x = K^*(x'',x',t''-t') K(x'',x',t''-t') d x
\nonumber
\\
\phantom{P(x'') d x}
= \left(1 - \frac{12 \beta m^2 (x'' - x')^2}{(t''-t')^2}\right)
\frac{m}{2 \pi \hbar (t''-t')} d x.
\label{probgup}
\end{gather}
It is clearly observable that the term $ \big(1 - \frac{12 \beta m^2 (x'' - x')^2}{(t''-t')^2}\big)$
in~\eqref{probgup} is smaller than 1 as $\beta > 0$.
Thus we conclude that the probability value in this case is less than the corresponding value in the free particle case.
Also, since probability is non-negative, this term $ \big(1 - \frac{12 \beta m^2 (x'' - x')^2}{(t''-t')^2}\big)$
should also be non-negative.
Thus we have the following relation
\begin{gather*}
1 - \frac{12 \beta m^2 (x'' - x')^2}{(t''-t')^2} \geq 0,
\end{gather*}
which immediately implies the bound for momentum as
\begin{gather}
p \leq p_{\max} = \frac{1}{2 \sqrt{3 \beta}},
\label{pmax1}
\end{gather}
where $p = \frac{m x}{t}$.
Thus, from~\eqref{pmax1} we see that GUP induces a~momentum upper bound in the theory which is comparable to maximum
momentum uncertainty~\eqref{pmax} induced by GUP.

Using this procedure, one can also calculate the GUP kernel for higher order terms of~$\beta$, i.e.~terms up to ${\cal
O}(\beta^2)$~\cite{pram2}.
It is interesting to note that the maximum momentum bound for~${\cal O}(\beta^2)$ case is less than that obtained in the
previous ${\cal O}(\beta)$ case, though by a~very small amount (please see~\cite{pram2} for details).

\section{Summary and discussion}\label{Section7}

A particular framework for quantum gravity is the doubly special relativity (DSR) formalism, that introduces a~new
observer independent energy scale, considered to be of the order of the Planck energy.
Presence of this energy scale naturally invokes noncommutative phase space structure.
We study the ef\/fects of this energy upper bound in relativistic thermodynamics for the particular case
of~$\kappa$-Minkowski spacetime where~$\kappa$ plays the role of energy upper bound.
We have explicitly computed the expression for the energy-momentum tensor of an ideal f\/luid in DSR
framework~\cite{dasdsr}.
In deriving the result we exploited the scheme of treating DSR as a~nonlinear representation of the Lorentz group in
special relativity.

We have also studied the modif\/ications in the thermodynamic properties of photon gas in this DSR scenario where we have
an invariant energy scale~\cite{dasph}.
We show both analytically and graphically that the density of states and the entropy in this DSR framework are less than
the corresponding quantities in Einstein's special relativity (SR) theory.
We stress that the Lorentz symmetry is not broken in this model.
But due do the presence of an invariant energy upper bound in this theory, microstates can avail energies only up to
a~f\/inite cut-of\/f whereas in SR theory, microstates can attain energies up to inf\/inity.
The internal energy is modif\/ied in case of the DSR model and as a~consequence the expression for the specif\/ic heat is
also modif\/ied.

Though highly non-trivial, one can similarly study the behaviour of an ideal gas or fermion gas in case of this DSR
model.
There might be some modif\/ications in the Fermi energy level which in turn can modify the Chandrasekhar mass limit for
the white dwarf stars.
Thus astrophysical phenomena in DSR framework is
another issue which remains
to be addressed.

\looseness=1
Further, as we have the expression for energy-momentum tensor, one can study the cosmo\-lo\-gi\-cal aspects of the DSR model
using the Friedmann equations.
But this requires idea about the geometry sector (precisely the metric $g_{\mu \nu}$ and hence Einstein tensor $G_{\mu
\nu}$) which is still unknown in the context of the DSR model.
This still remains another open issue to be further studied.

It is noteworthy to mention here that ``bouncing'' loop quantum cosmology theories (for example see~\cite{bojowald} and
references therein) entail some modif\/ications to the geometry of spacetime which in turn ef\/fectively put a~bound on
the curvature avoiding the big bang singularity.
However, for these ``bouncing'' models, the perturbation technique cannot be used as at the point of curvature
saturation, the energy density of the cosmic f\/luid diverges.
So it is unclear how to construct the matter part of the Einstein equation.
One alternative to avoid the big bang singularity is the inf\/lation theory where the perturbation method can also be
applied.
On the other hand, in our model, the energy density of the cosmic f\/luid saturates to the Planck energy which is a~f\/inite
real quantity.
Possibly a~combination of the model considered in this paper along with the ``bouncing'' loop quantum cosmology can
successfully describe a~situation where big bang singularity can be avoided.

We also have considered generalized uncertainty principle (GUP) induced models~\cite{pram1}.
We derived a~covariant free particle Lagrangian in GUP and also studied GUP Lagrangian in pre\-sen\-ce of an external
electromagnetic f\/ield.
We show how the equations of motion are af\/fected by the noncommutative parameter present in the theory.

GUP gives rise additional terms in quantum mechanical Hamiltonian like $\beta p^{4}$, where
$\beta\sim\frac{1}{(M_{\rm P}c)^{2}}$ is the GUP parameter.
This term plays important role at Planck energy level.
Consi\-de\-ring this term as a~perturbation, we have shown that path integral method is applicable on this GUP corrected
non-relativistic cases~\cite{pram2}.
Here we have constructed the explicit form of the GUP kernel by applying Hamiltonian path integral method.
The consistency properties of this kernel is then thoroughly verif\/ied.
We have shown that the probability for f\/inding a~particle at a~given point in case of the GUP model is less than the
corresponding probability in the canonical free particle case.
We have also shown that probabilistic interpretation of this kernel induces a~momentum upper bound in the theory.
And this upper bound changes slightly with $e^{-{\cal O}(\beta^2)}$, if we consider higher order~$\beta$ terms in the
Hamiltonian.
Following this Hamiltonian path integral approach one can construct kernels and study their properties for other systems
like particle in a~step potential, Hydrogen atom etc, which is our future goal.

\pdfbookmark[1]{References}{ref}
\LastPageEnding

\end{document}